\documentclass[twocolumn, superscriptaddress,showpacs,amsmath,amssymb,prl]{revtex4-1}
\usepackage{graphicx}
\usepackage{dcolumn}
\usepackage{bm}

\usepackage{times,mathptmx}
\usepackage{txfonts}
\usepackage{xcolor}
\usepackage{natbib}
\usepackage[hidelinks]{hyperref}
\hypersetup{
	colorlinks   = true, 
	urlcolor     = black, 
	linkcolor    = red, 
	citecolor   = blue 
}

\DeclareFontFamily{OMS}{oasy}{\skewchar\font48 }
\DeclareFontShape{OMS}{oasy}{m}{n}{%
	<-5.5> oasy5     <5.5-6.5> oasy6
	<6.5-7.5> oasy7     <7.5-8.5> oasy8
	<8.5-9.5> oasy9     <9.5->  oasy10
}{}
\DeclareFontShape{OMS}{oasy}{b}{n}{%
	<-6> oabsy5
	<6-8> oabsy7
	<8->  oabsy10
}{}
\DeclareSymbolFont{oasy}{OMS}{oasy}{m}{n}
\SetSymbolFont{oasy}{bold}{OMS}{oasy}{b}{n}

\DeclareMathSymbol{\smallleftarrow}     {\mathrel}{oasy}{"20}
\DeclareMathSymbol{\smallrightarrow}    {\mathrel}{oasy}{"21}
\DeclareMathSymbol{\smallleftrightarrow}{\mathrel}{oasy}{"24}

\newcommand{\monote}[1]{\textsf{{\color{blue} \scriptsize 
	}}\marginpar{{\textbf{Note}}}}

\begin{document}
	
	\title{Observation of photonic spin-momentum locking due to coupling of achiral metamaterials and quantum dots}

	\author{Ravindra Kumar Yadav}
	\affiliation{Department of Physics, Indian Institute of Science, Bangalore 560012, India}
	\author{Wenxiao Liu}
	\affiliation{Texas A\&M University, College Station, Texas 77843, USA}
	\affiliation{Shaanxi Province Key Laboratory for Quantum Information and Quantum Optoelectronic Devices, Xi$^{\prime}$an Jiaotong University, Xi$^{\prime}$an, 710049, China}
	\author{SRK Chaitanya Indukuri}
	\affiliation{Department of Physics, Indian Institute of Science, Bangalore 560012, India}
	
	\author{Adarsh B. Vasista}
	\affiliation{Department of Physics, Indian Institute of Science Education and Research(IISER) Pune-411008, India}
	\author{G. V. Pavan Kumar}
	\affiliation{Department of Physics, Indian Institute of Science Education and Research(IISER) Pune-411008, India}
	\author{Girish S. Agarwal}
	\email{girish.agarwal@tamu.edu}
	\affiliation{Department of Physics, Indian Institute of Science, Bangalore 560012, India}
	\affiliation{Texas A\&M University, College Station, Texas 77843, USA}
	\altaffiliation{Department of Physics, Indian Institute of Science, Bangalore 560012, India}
	\email{girish.agarwal@tamu.edu}
	\author{Jaydeep K Basu}
	\email{basu@iisc.ac.in}
	\affiliation{Department of Physics, Indian Institute of Science, Bangalore 560012, India}
	
	\date{\today}
	
	\begin{abstract}
		Here, we report observations of photonic spin-momentum locking in the form of directional and chiral emission from achiral quantum dots (QDs) evanescently coupled to achiral hyperbolic metamaterials (HMM). Efficient coupling between QDs and the metamaterial leads to emergence of these photonic topological modes which can be detected in the far field. We provide theoretical explanation for the emergence of spin-momentum locking through rigorous modeling based on photon Green's function where pseudo spin of light arises from coupling of QDs to evanescent modes of HMM.
		
	\end{abstract}

	\maketitle
	Topological photonics\cite{ozawa2019topological} is an emerging area where various novel phenomenon like 
	photonic spin Hall effect\cite{bliokh2015quantum,bliokh2014extraordinary,bliokh2015spin,van2016universal}, photonic topological insulators\cite{khanikaev2013photonic,rider2019perspective,lu2014topological}, unidirectional propagation of light\cite{petersen2014chiral,le2015nanophotonic,sollner2015deterministic,rodriguez2013near,o2014spin} etc.
	have been reported using various degrees of freedom of light coupled
	with advanced design of metamaterial and metasurfaces\cite{high2015visible,tan2014photonic}. A key feature of topological effects, spin-momentum locking has been reported for photonic systems where extrinsic helicity of light or explicitly chiral emitters are coupled to  surface plasmon polaritons(SPP) or waveguide modes\cite{rodriguez2013near,mitsch2014quantum} in near field\cite{kapitanova2014photonic}.Various chiral structures have been used to demonstrate chirality in emission including nanofibers with trapped nanoparticles or atoms on it \cite{le2006angular, petersen2014chiral,mitsch2014quantum} and engineered photonic crystal waveguides having chiral modes \cite{sollner2015deterministic,le2015nanophotonic,young2015polarization,hughes2017anisotropy}.
	 Another system which has been widely used to demonstrate photonic topological effects are metamaterials which have several fascinating properties \cite{smith2004metamaterials,fang2009optical,sreekanth2013directional,zhang2009negative,soskin2016singular,zhang2009negative,gao2015topological}, etc.
	
		A special class of metamaterials which possesses a hyperbolic iso-frequency surface known as hyperbolic metamaterials (HMM) have a number of novel properties like anisotropy and large photonic density of states which can be used for a number of applications \cite{poddubny2013hyperbolic,lu2014enhancing,biehs2018strong,biehs2016long,starko2015optical,berry2003optical,krishnamoorthy2012topological}. 
The high wavevectors (high-$k$) of an emitter are evanescent in conventional dielectric media but HMM supports propagation of high-$k$ wavevectors \cite{west2015adiabatically,cortes2012quantum,jahani2018controlling}. However,  it has not yet been possible to utilize the properties of these modes by transporting them to the far-field. In general, it has been demonstrated that when chiral emitters couple with evanescent high--$k$ modes of HMMs unidirectional and chiral emission emerges in the  near-field\cite{guddala2019optical,kapitanova2014photonic}. 
	HMM provides the extrinsic momentum to photon by making it directional which couples to intrinsic spin of photon (polarisation state of light) analogous to a spin-orbit interaction.
 However, transport of these photonic topological modes, especially with intrinsically achiral structures, to the far field has not yet been reported.
	
	Here, we report experimental observation and theoretical modeling of photonic topological effects in the form of  directional and chiral far field emission from intrinsically achiral semiconductor quantum dots (QDs) coupled to a two dimensional (2D) achiral HMM. We observe clear signatures of directional and chiral emmission from the hybrid  system in wavevector resolved emission map in experiments. We report a new regime of spin-momentum locking phenomenon where extrinsic pseudo helicity of light is provided by coupling quantum emitters to evanescent high-$k$ modes of HMM. The observed directional and chiral emission, in experiments, coincides with the appearance of strong splitting of the photoluminescence (PL) spectra from the QDs on HMM\cite{indukuri2017broadband, SM}. The observed experimental results are explained reasonably well using rigorous numerical scheme to evaluate electric field distribution using method of photon Green's function for geometry similar to that used in experiment. Significantly, we also demonstrate the ability to detect the photonic topological modes generated due to the high-$k$ evanescent HMM modes to the far-field. Overall, our  study provides a methodology to obtain far field photonic spin-momentum locking without having to use chiral emitters or chiral metamaterials.
	
	\begin{figure}
		\centering
		\includegraphics[scale=0.44]{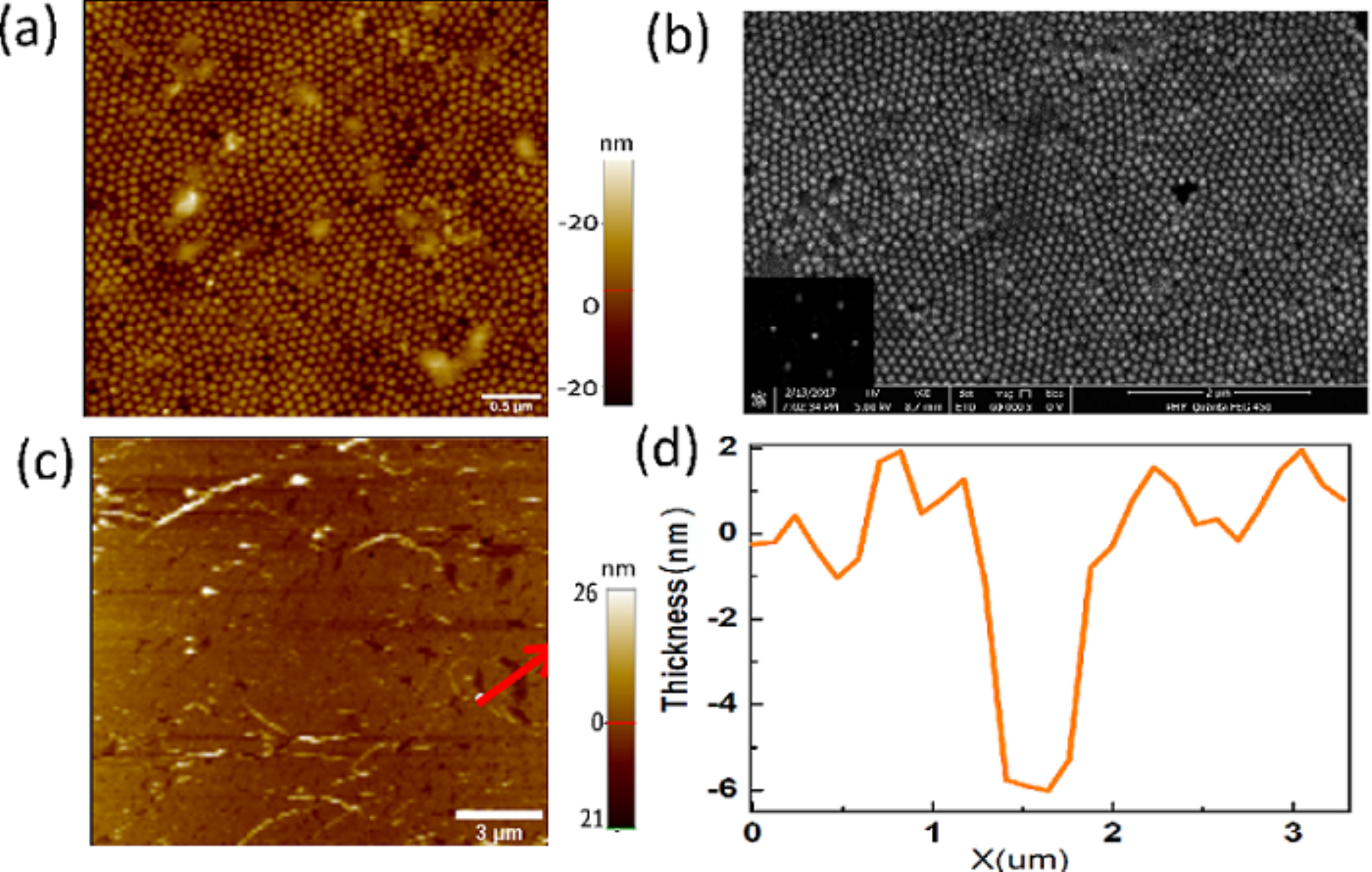}
		\caption{\label{cqed_figure}
(a) AFM image of top surface of HMM template before transfer of QD monolayer. (b) SEM image of top surface of HMM template showing silver nanowire diameter of $\sim$50nm and inter-wire sepration of $\sim$ 110nm (Inset shows fast Fourier transformation (FFT) of SEM image). (c) shows the AFM image of HMM after transfer of compact QD monolayer with the corresponding line profile along the red line shown in (d) }     
	\end{figure}
	
	The experimental results are based on fabricated 2D HMM templates consisting of hexagonal arrays of silver nanowires (length $800~\text{nm}$) embedded in alumina matrix prepared using methods described earlier \cite{kanungo2010experimental}. A compact monolayer of cadmium selenide (CdSe) QDs, synthesised by well known methods\cite{peng2001formation,indukuri2017broadband,praveena2015plasmon}, was placed on top of the metamaterial using the well known Langmuir-Blodgett (LB) technique as discussed earlier \cite{dabbousi1994langmuir,haridas2011photoluminescence} as well as in supplementary materials \cite{SM}. The metamaterials prepared undergo a wavelength dependent topological transition to a phase having hyperbolic optical dispersion - hence called HMM - depending on silver nanowire filling fraction ($f$ = 0.15 in our study) \cite {SM}. The peak emission wavelength of the QDs was chosen such that it lies well inside HMM dispersion spectral regime \cite {SM}.
Atomic force microscopy (AFM) and scanning electron microscopy (SEM) image of top surface of HMM is shown in Fig.~1(a) and (b), respectively, reveals the ordered structure of the array as well as the lattice spacing. AFM image of the same template after transfer of a compact QD monolayer on top of a polymer spacer is shown in Fig 1(c) with the thickness of the film ($7~\text{nm}$) being revealed in a typical height profile in Fig 1(d) roughly correponding to the diameter of the QDs used \cite {SM}. For back focal plane Fourier imaging the QD monolayers on HMM templates were excited  by focusing a 633 nm continuous wave (CW) laser using 100X objective with numerical aperture (NA) being 0.95. Emission was collected by the same objective\cite{vasista2018differential}. For PL spectro-microcopy measurements excitation was performed with 514nm laser as described earlier \cite{indukuri2017broadband} and in \cite {SM}.

	Fig.~2(a) shows schematic of experimental system which demostrates the directional and chiral emission from QD monolayer coupled to HMM.  In Fig.~2(b-c), we have shown the Fourier images for QDs on HMM and QD on glass, respectively for spacer layer thickness, $d$ = 10nm.
	\begin{figure}
		\centering
		\includegraphics[scale=0.45]{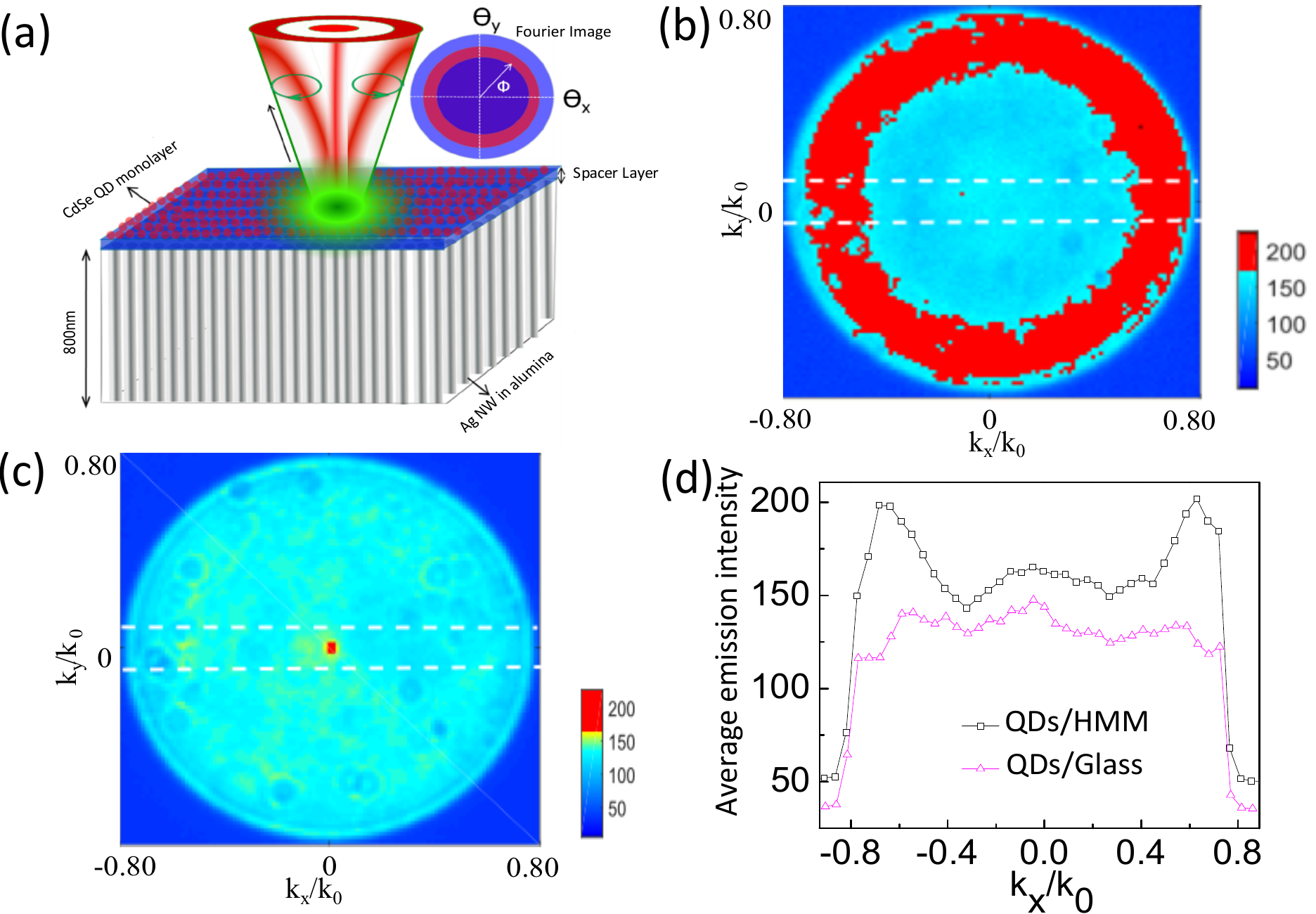}
		\caption{\label{cqed_figure}
(a)Schematic of the experimental setup while the inset shows a typical back focal plane Fourier image pattern. Fourier space emission profiles with unpolarised excitation for (b) QDs on HMM corresponding to $d=10~\text{nm}$ and (c)for QD on glass. (d)QD emission intensity as a function of in-plane wave vector $k_{x}/k_{0}$ for the patterns shown in (b) and (c). }     
	\end{figure} 
	A clear signature of directional emission was observed for QDs placed in the near-field of HMM and emission is azimuthally symmetric and wavevector preferential. Azimuthally symmetric rings of maximum intensity are observed for QDs placed in near-field of HMM at a separation of 10 nm, which is shown in Fig.~2(b). We also performed a reference measurement for same QD monolayer on glass as shown in Fig. 2(c). Figure 2(c) shows a completely isotropic and homogeneous emission from the same QD monolayer on glass. To quantify this data further, we extracted the line profile of emission intensity for a band of $k_{y}$ at $k_{y}/k_{0}=0\pm 0.12$ along $k_{x}$ as indicated in Fig.~2(b)and(c) by a pair of lines. Emission intensity is integrated over a band of $k_{y}/k_{0}=0\pm 0.12$ to avoid the spatial resolution limited fluctuations in emission intensity. These integrated intensity profiles when plotted as a function $k_{x}/k_{0}$ in Fig.~2(d) reveal two symmetric peaks of maximum emission intensity $k_{y}/k_{0}=\pm 0.70$ for $d$ =10nm, whereas for the reference sample (QD on glass) it does not show any wavevector preferential emission. In summary, Fig.~2 manifests that the emergence of directional emission from QDs on HMM is due to the coupling of excitons in QDs to large-$k$ modes of HMM.
	
	After confirmation of the connection of directional emission with large-$k$ modes, we explore the study of the chiral nature of these modes. In general, these large-$k$ modes are evanescent in nature. To study this,we excited our system with  LCP and RCP polarised $633~\text{nm}$ CW laser and emission was collected in the Fourier image. Fig.~3 shows LCP and RCP excitation dependent Fourier images for QDs on HMM and QDs on glass systems. It should be noted that we have eliminated the central region( NA= 0 to $\pm$ 0.24) of the Fourier image which has emitted signal of the uncoupled QDs with large-$k$ modes of HMM. The resultant Fourier image of LCP and RCP excitation shows a circular ring of maximum intensity followed by a minima located at mutually opposite azimuthal angles ($\phi$) as visible in Fig.~3(a) and (b), which shows the helicity dependent emission. The radius of the ring gives us the directions of emission which are characterized by $k_{x}$ and $k_{y}$. 
	\begin{figure}
		\centering
		\includegraphics[scale=0.15]{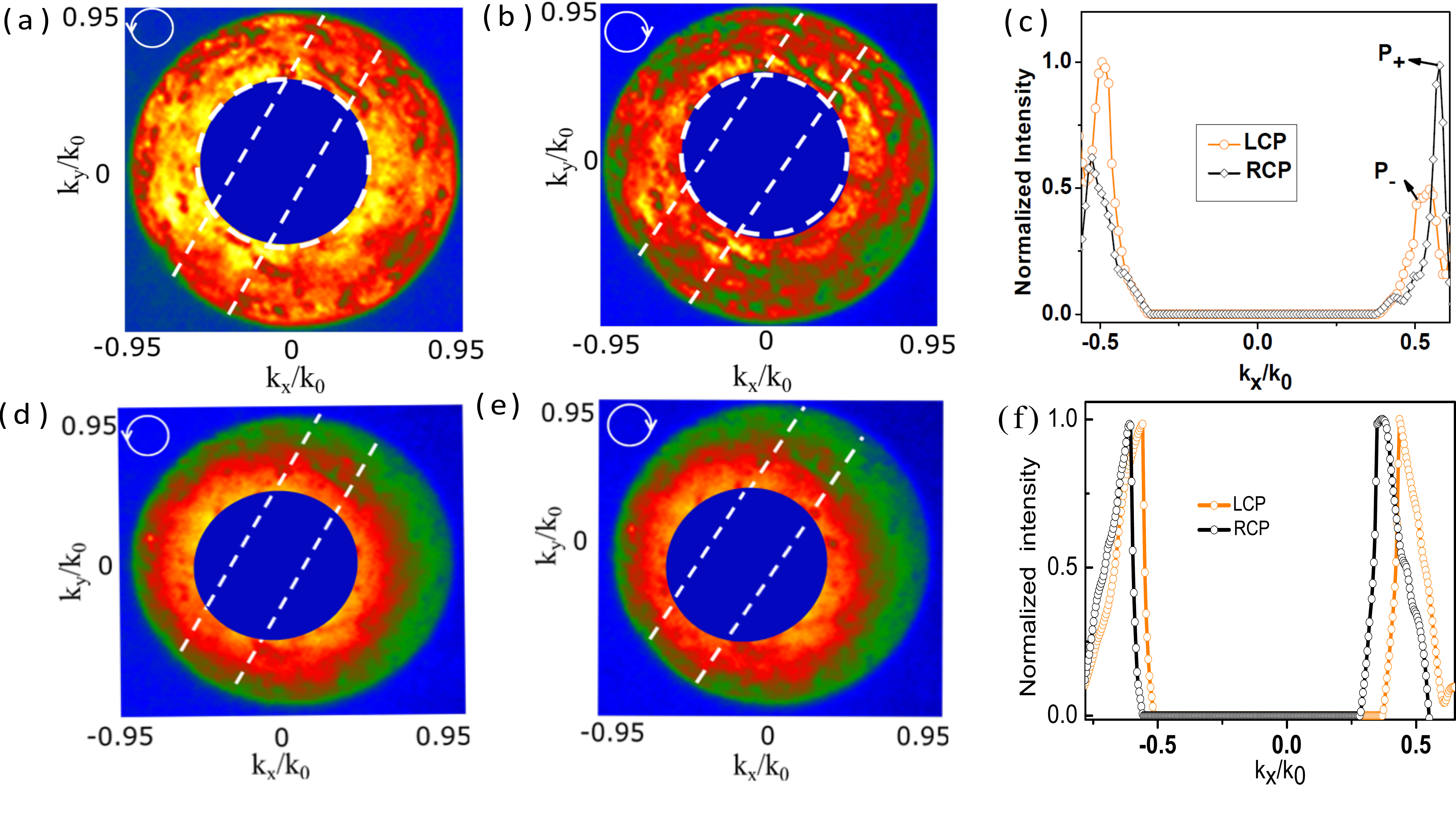}
		\caption{\label{cqed_figure}
Fourier image corresponding to $d$=10nm as function of incident helicity for QDs on HMM. (a) LCP (b) RCP. (c) Intensity profile normalised with maximum intensity in each case as function of wavevector along polarization direction and ${P_+}$, ${P_+}$ are intensity magnitudes of LCP
			and RCP excitation dependent emission at ${k_x}/k_{0}=0.50$. (d) LCP for QDs on glass, (e) RCP for
			QDs on Glass. (f) Shows emission intensity profile as function of wavevector along the polarization
			direction.}    
	\end{figure} 
To extract the emission intensity line profiles from the Fourier images, a pair of white dotted lines are drawn in a fixed direction. Extracted emission intensity profile is normalised with maximum intensity of extracted line profile and plotted in Fig.~3(c). Normalised emission intensity profile shows antisymmetric intensity peak for LCP and RCP excitations. For RCP excitation, maximum intensity peak occurs at $k_{x}>k_0$ and the case is reverse for LCP excitation. After confirming the chiral nature of emission from HMM based system, we calculated chirality ($C$) using following expression: 
		\begin{equation} 
		C=\frac{I_{m}(P_{+})-I_{m}(P_{-})}{I_{m}(P_{+})+I_{m}(P_{-})},
		\end{equation}
		where $I_{m}(P_{-})$ and $I_{m}(P_{+})$ are local maximum intensity for $k_{x}>k_{0}$ for LCP and RCP
		incident polarizations and we have $0 \leq C \leq 1$. The value $C=0$ stands for the excitation by using a linear polarization, while $C=1$ implies a truly unidirectional excitation of the field. The chirality parameter was obtained $C=0.35$ from above mentioned analysis of Fourier image as indicated in Fig.~3(c). We also performed LCP and RCP excitation dependent Fourier imaging for QDs on glass as reference measurement as shown in Fig.~3(d) and 3(e), respectively. Emission intensity line profiles for QDs on glass are plotted in Fig.~3(f) which clearly shows no chiral pattern. Thus, our experimental observations suggest the ability of the QDs to couple HMM high $k$ modes leading  to observation of spin-momentum locking detectable in far-field. Interestingly , this phenomenon disappears for larger value of spacer thickness,$d$, ( \cite{SM}, Figure S4), suggesting  an intricate connection between the emergence of spin-momentum locking and the observation of splitting of PL spectra for QDs on HMM templates as shown in detail by us earlier \cite{indukuri2017broadband} and in \cite{SM}(Figure S3).
		
		\section*{Theoretical Modeling}
	
		We next develop a model for the observations. We use rigorous solutions of Maxwell equations, i.e., the development is similar to a recent one \cite{liu2019chiral} but is generalized to the present experimental geometry. Unlike the layered medium in the previous work, the HMM in the present work is different and is characterized by complex effective dielectric tensor. The detailed modeling is given in the Supplementary Material \cite{SM}. We can numerically investigate the emission characteristics radiating by a two-dimensional dipole on the nanowire HMMs, where the unit dipole moment of the dipole is $ \bm{p}_\text{0}=p_{x} \hat{\bm{x}}+p_{z} \hat{\bm{z}}$ with $p_{x}^{2}+p_{z}^2=1$. We select the related parameters corresponding to the experiment as: $\epsilon_{1}=\epsilon_{4}=1$, $\epsilon_{2}=2.5$, $d=10~\text{nm}$, and $l=800~\text{nm}$. In order to make the coupling mechanism clear, we first consider the case of $k_{y}=0$. In this case, the s-polarized waves have no contribution to the electric field intensity from the Eq.~(S8)\cite{SM}. Then the electric field intensity $I^{p}$ of the p-polarized waves in the Fourier transformed space can be described as \cite{novotny2012principles}
		\begin{equation} 
		I^{p}=\frac{|r^{p}|^2(k_{x}^2+|k_{1z}|^2)}{k_{0}^2} \cdot|p_{x}+p_{z}\frac{k_{x}}{k_{1z}}|^{2},
		\end{equation}
		where $k_{1z}=\sqrt{k_{0}^2-k_{x}^2}$ and $k_{0}=\omega/c$. As discussed in \cite{SM}, the reflection $|r^{p}|^2$ is symmetric along the $k_{x}=0$. Therefore, the Eq.~(2) can be seemed as $I^{p} \propto |p_{x}+p_{z}k_{x}/k_{1z}|^{2}$. For evanescent waves with $|k_{x}|>k_{0}$, the term $k_{1z}$ is a complex number and then $I^{p} \propto |p_{x}-i p_{z}k_{x}/|k_{1z}||^{2}$. If we let $p_{x}$ and $p_{z}$ be real numbers without a $\pi/2$ phase difference used for the unpolarized case, the electric field intensity can be written as $I^{p}\propto |p_{x}|^2+ |p_{z}k_{x}/k_{1z}|^{2}$, which is a linear superposition of the contributions from the horizonal and vertical oriented components of the dipole. No interference phenomenon occurs in this case. However, the case is different when $p_{x}$ and $p_{z}$ have a $\pi/2$ phase difference. For instance, for the right-circularly polarized dipoles, we have $p_{x}=\frac{1}{\sqrt{2}}$ and $p_{z}=\frac{i}{\sqrt{2}}$. Consequently, the electric field intensity can be reduced to $I^{p} \propto |1+k_{x}/|k_{1z}||^{2}/2$. 
		The horizonal and vertical oriented spectral amplitudes for $k_{x}>k_{0}$ add up constructively, whereas for $k_{x}<-k_{0}$ destructive interference occurs. Therefore, we emphasize that chiral emission occurs when a circularly polarized dipole excites the multilayered nanostructure. More generally, elliptic polarization with a phase difference of $\pi/2$ between the $x$ and $z$ components is required. These very general findings on the expected chiral behaviour are in agreement with the experimental results in Fig.~(3). We also disccused the case without absorption in silver, which is given in the SM( figure S7))\cite{SM}.
				
		\begin{figure}
			\centering
			\includegraphics[scale=0.40]{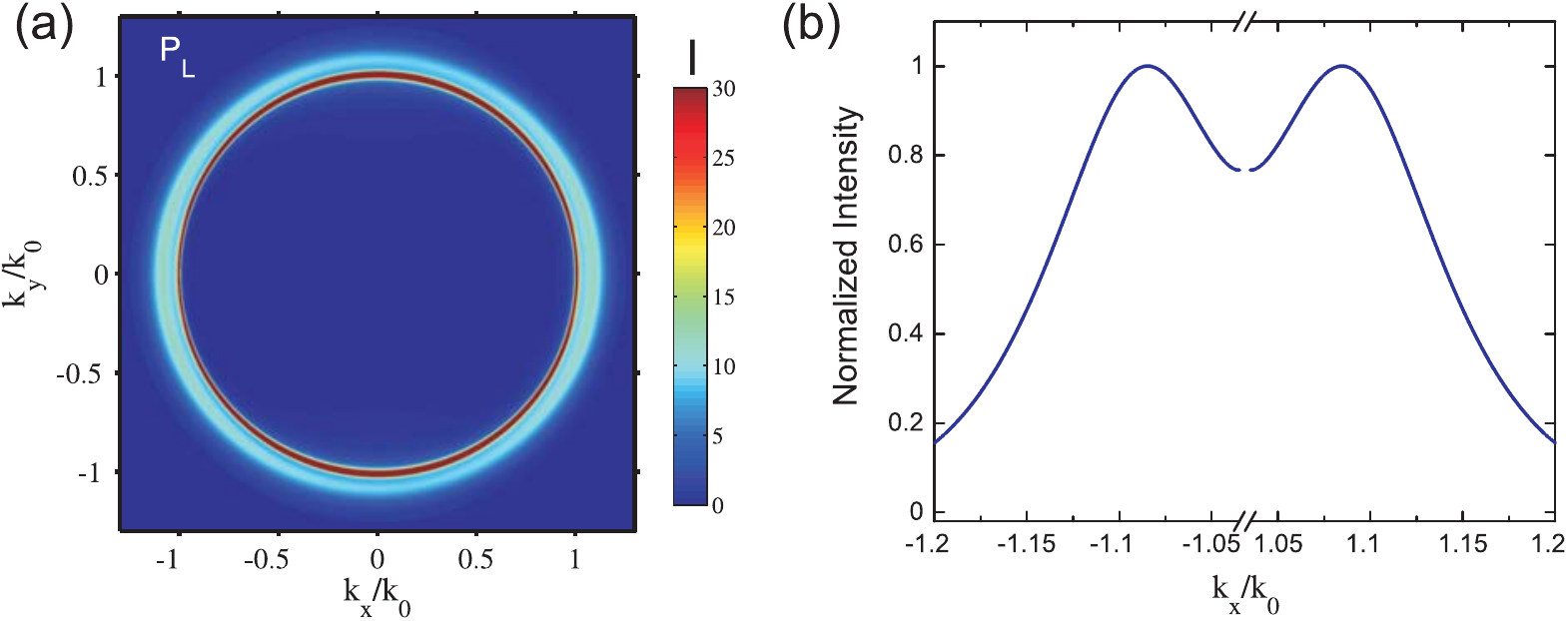}
			\caption{(Color online) (a) Electric field intensity map of QDs/ HMM for unpolarised excitation with $p_{x}=1/\sqrt{2}$ and $p_{z}=1/\sqrt{2}$. (b) Emission intensity profile as a function of $k_{x}$ for the case of $k_{y}=0$.}
		\end{figure}
			
		In the following, we give a quantitative analysis about the electric field intensity ($I^{p}$) for different dipolar polarization states. The unit vector of the horizontally (vertically)  polarized dipole is $P_{x}=\hat{\bm{x}}$ ($P_{z}=\hat{\bm{z}}$). For the circularly polarized dipoles, the unit polarization vectors are $P_{\pm}=\frac{1}{\sqrt{2}}(\hat{\bm{x}}\pm i \hat{\bm{z}})$. The dielectric constant tensor for the HMMs can be described by the effective medium theory, as given in \cite {SM}, by 
$\epsilon_{\parallel}=3.79+0.0088i$ and $\epsilon_{\perp}=-0.843+0.0672i$ for the wavelength $\lambda=660~\text{nm}$.In Fig.~4(a), we  present the electric field intensity map induced by the unpolarized dipole with $p_{x}=1/\sqrt{2}$ and  $p_{z}=1/\sqrt{2}$, which can match well with that in Fig.~2(b) and 2(d). Figure 4(b) shows the normalized emission intensity profile along the $k_{x}$ direction with $k_{y}=0$. It has two symmetric peaks with approximately equal intensity at $k_{x}/k_{0}=\pm 1.085$, which denotes that there is no chiral behaviour but only directional characteristics. 

		 The intensity maps containing information for both TE and TM modes for different dipolar polarizations are shown in Fig.~5(a-b) and 5(d-e), which give a circular ring of maximum intensity followed by a minima and show directional emission. Moreover, chiral emission for the cases of circularly polarized dipoles with $\textbf{P}_{\pm}$ can be found in Fig.~5(a) and 5(b). The electric field profile is asymmetric along the $k_{x}$ axis in Fig.~5(a-c). According to the Eq.~(2), the incident field induced by a left-circularly polarized dipole ($P_{+}$) couples more efficiently with a forward-propagating mode for $k_{x}>k_{0}$, and backward propagation is suppressed for $k_{x}<-k_{0}$. However, the case for a right-circularly polarized dipole ($P_{-}$) is converse. For the right-circularly polarized dipole, the position of peak appears at $k_{x}=1.088k_{0}$ and  the peak for $k_{x}<-k_{0}$ appears as a relatively broad shoulder. Even though, emission intensity line profiles still show chiral behaviour for circularly polarized dipoles with $C=0.682$ at $k_{x}=1.088k_{0}$. This theoretical value of $C$ is higher than the experimental value. This is because the theoretical result is calculated from the peak value, whereas the experiment data is averaged over a range of $k_{x}$ vectors. This clearly will bring down the observed value of $C$.  Instead, for linearly polarized dipoles, the excited field is bidirectional propagation, as shown in Fig.~4(f). The forward-propagating and backward-propagating modes are symmetrical modal distribution. Electric field intensity of the high-$k$ propagating mode induced by a z-polarized dipole is larger than that of an x-polarized dipole. We can find that even though there is no preference in the $k_{y}$ direction, the directional emission can be realized by using the circularly polarized dipoles.In the experimental data, the measured wave vectors are in free space where as in theoretical modeling $k_{x}$ and $k_{y}$ are in glass objective which has refractive index 1.45. A closer look at Fig.~3(a) shows for example that the emission max-min corresponds to $\kappa$ about 0.7 which translates to $\kappa$ in glass as 1.015 which matches well with the maximum in the Fig.~5(c) obtained in the electromagnetic calculations.

	\begin{figure}
			\centering
			\includegraphics[scale=0.40]{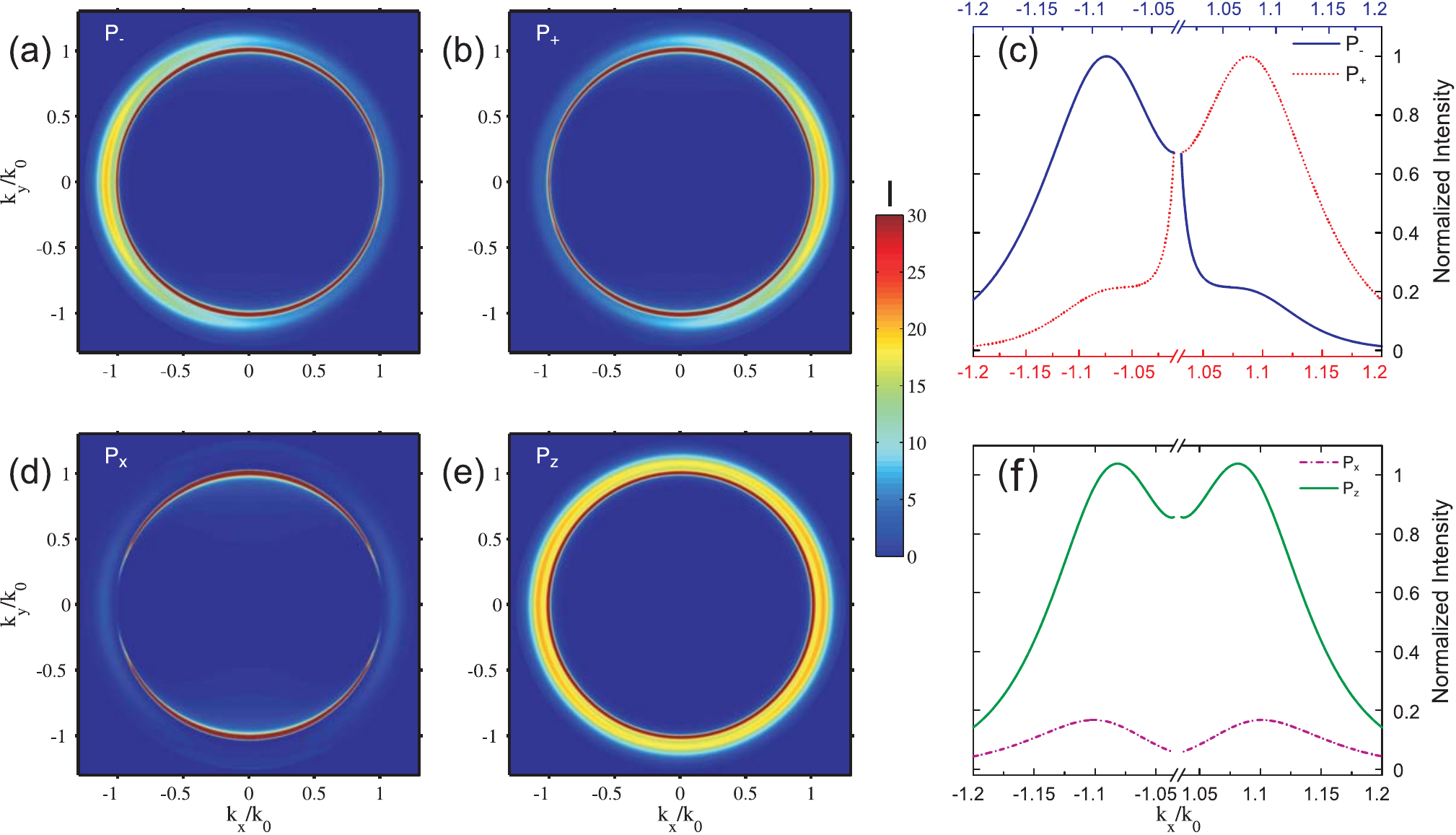}
			\caption{(Color online) Electric field intensity maps when considering the absorption in sliver with different polarization, (a) $ \textbf{P}_\text{-}=\frac{1}{\sqrt{2}}( \hat{\textbf{x}}-i\hat{\textbf{z}})$, (b) $ \textbf{P}_\text{+}=\frac{1}{\sqrt{2}}( \hat{\textbf{x}}+i\hat{\textbf{z}})$, (d) $ \textbf{P}_\text{x}= \hat{\textbf{x}}$, and (e)$\textbf{P}_\text{z}= \hat{\textbf{z}}$.
				Intensity profile induced by circularly polarized (c) and linearly polarized (f) dipoles as a function of $k_{x}$ are plotted for $k_{y}=0$. The electric field $E_{c}$ used for normalization corresponds to the larger peak value induced by a circularly polarized dipole in Fig.~5(c).}
		\end{figure}
	The polarization degree of freedom provides a new approach for studying a variety of novel optical effects. Finally, we show how the dielectric constant tensor $\epsilon_{\parallel}$ and $\epsilon_{\perp}$ can be used to obtain the chiral character in a simple manner. 
		\begin{figure}
			\centering
			\includegraphics[scale=0.5]{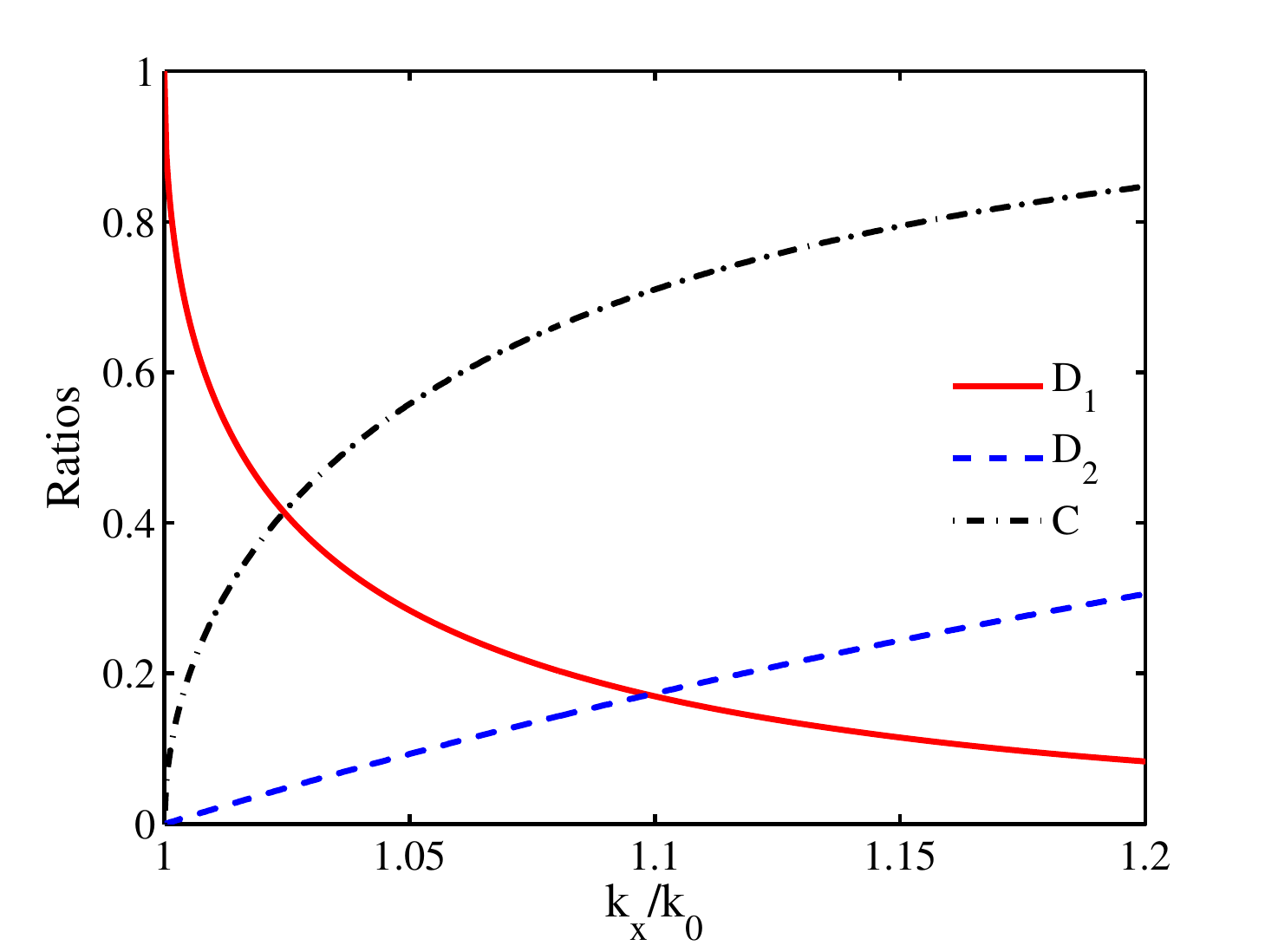}
			\caption{(Color online) The intensity ratios ($D_{1}$ and $D_{2}$) and chiral parameter ($C$) as a function of the wave vector component $k_{x}$. Here, $k_{y}=0$ and $k_{x}>k_{0}$.}
		\end{figure}
		For large-$k$ wave vectors ($k_{x}>k_{0}$), the intensity ratio between the left-circularly ($P_{-}$) and right-circularly ($P_{+}$) polarized components is defined as
		\begin{equation} 
		D_{1}=\frac{I^{p}(P_{-})}{I^{p}(P_{+})}=(\gamma-\sqrt{\gamma^2-1})^4,
		\end{equation}
		where $\gamma=k_{x}/k_{0}$. The chiral parameter $C$ can be given by $C=(1-D_{1})/(1+D_{1})$. We can also define the intensity ratio induced by the horizontally and vertically polarized dipoles as
		\begin{equation} 
		D_{2}=\frac{I^{p}(P_{x})}{I^{p}(P_{z})}=1-1/\gamma^2.
		\end{equation}
		In Fig.~6, we present the intensity ratios ($D_{1}$ and $D_{2}$) and chiral parameter ($C$) as a function of the wave vector component $k_{x}$ for $k_{x}>k_{0}$ and $k_{y}=0$. When increasing the value of $k_{x}$, the intensity ratio $D_{1}$ goes down, while the chiral parameter $C$ and intensity ratio $D_{2}$ grow up. Therefore, we can obtain larger chiral value by properly changing the filling ratio and the emission wavelength.   Thus we can estimate the chirality depending on the direction in which emission is dominant. This value of $k_{x}$ is determined by the properties of the hyperbolic medium and Eq.3 implies the spin momentum locking which we observe in our experiments. 
				In conclusion, we have demonstrated photonic spin-momentum locking in the form of directional and chiral emission from excitons in compact monolayers of QDs efficiently coupled to evanescent metamaterial modes in a 2D HMM. Based on the method of photon Green's function, the chiral emission from a circularly polarized dipole coupled with HMMs are presented theoretically and the origin of photonic spin-momentum locking in experiments, quantitatively, explained. The origin of chirality is attributed to the near-field interference between the longitudinal and transverse polarized components of the dipolar QDs in evanescent high-$k$ modes of the HMM. Equivalently, the directional emission can be observed by exploring a elliptically polarized dipole to break the inversion symmetry, similar to the spin Hall effect.  Our  study thus provides a method to obtain chiral and directional photonic effects in the far field without having to use chiral emitters or chiral metamaterials.

		\begin{acknowledgments}
			The authors acknowledge Indo-U.S. Science and Technol-ogy Forum (IUSSTF) for funding through a virtual centre onquantum plasmonics.The authors also acknowledge Scheme for Promotion of academic and Research Collabration(SPARC) for funding.WXL is supported by the National Natural Science Foundation of China (NSFC) (Nos.~11534008, 91536115), and the Natural Science Foundation of Shaanxi Province (No.~2016JM1005). GSA thanks the support from the Welch Foundation (No.~A-1943-20180324). GSA also thanks the Infosys Foundation Chair of  Department of Physics, IISc Bangalore which made this collaboration successful.GVPK acknowledges the support of Swarnajayanthi fellowship grant - DST/SJF/PSA-02/2017-18 and Center for Energy Science. Grant Number: SR/NM/TP‐13/2016.
		\end{acknowledgments}

		\clearpage
		
	\end{document}


\newpage
	\section{Synthesis of CdSe QDs}
	For Topo caped CdSe core synthesis, Cadmium oxide (6.6mg) , Hexaphasphonic
	acid (40mg) and tryoctylephosphine oxide (TOPO) (1.85gram) was taken in three neck
	flask (25ml capacity). Selenium (Se) precursor was prepared by dissolving 20.5mg
	Se in 1gram tryoctylephosphene (TOP) in cleaned voil using mild shaking. Finally,
	three neck flask mounted on heating mental as and heated up to 2600c under nitrogen
	gas (N2) flow to get colourless solution while heating set temperature was kept at 320oC
	and current 5mA. Se precursor was added quickly using syringe in three neck flask
	at 320 oc and kept for 20min to get QDs with emission wavelength 660nm and system
	was kept cooled down to room temperature . Reaction was quenched by adding 2 ml
	toluene in three neck flask at room temperature. QDs was cleaned by centrifuge the
	1:3 mixture of stock QDs solution in toluene: methanol with speed 12000rpm for 10min
	for three time . Finally QDs was redispersed in chloroform for monolayer transfer using
	LS method. Fig S1 shows the Solution photoluminescence and absorption of CdSe QDs in toluene.\cite{indukuri2017broadband}.
	\begin{figure}
		\centering
		\includegraphics[scale=0.35]{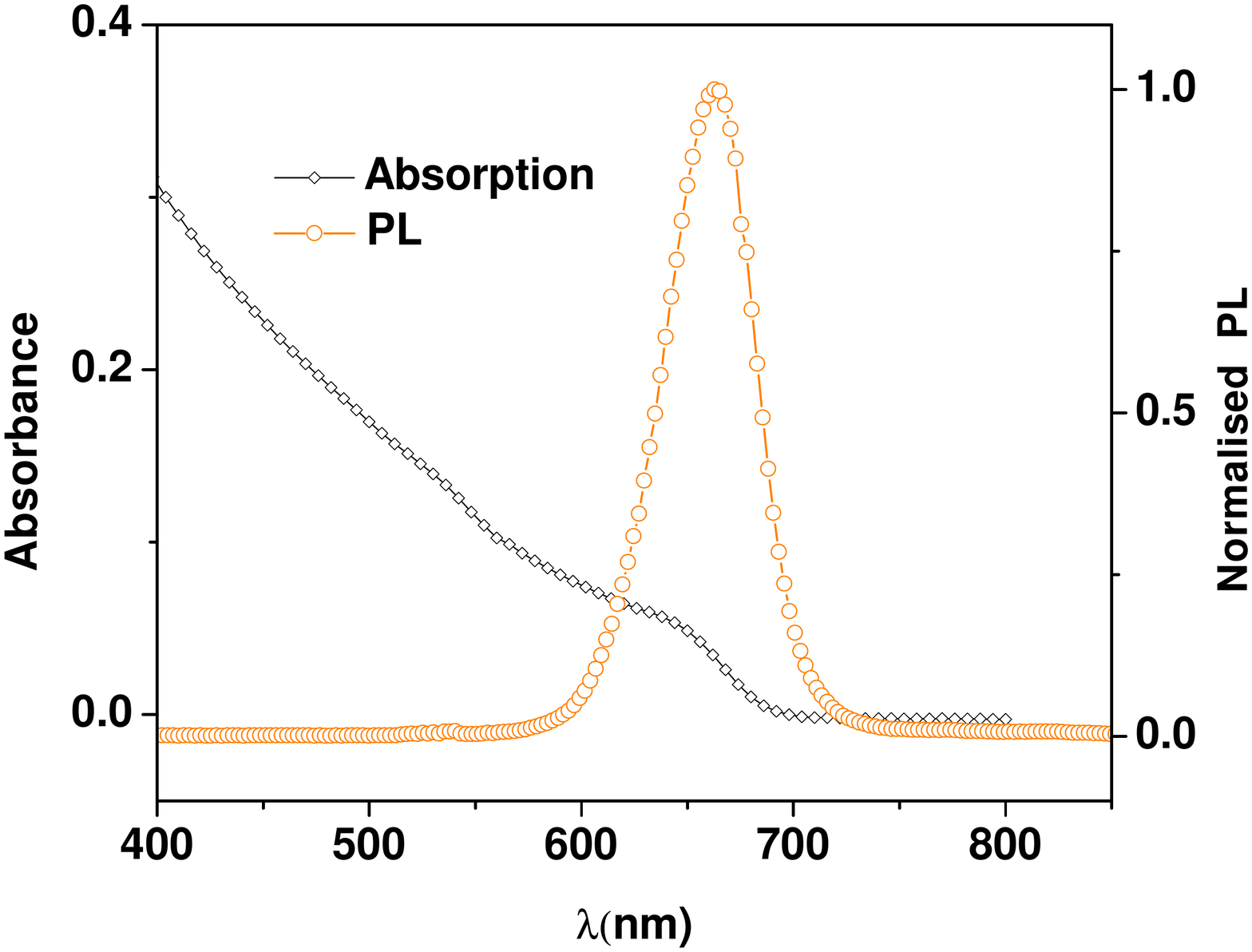}
		\caption{Absorption and photoluminescence of CdSe QDs}
		\label{fig:fl_map}
	\end{figure}
	
	\section{Fabrication of HMM}
	Highly pure aluminium sheet (99:9999$\%$) were purchased from Alpha Asser. Aluminium
	sheets were sonicated in ethanol for two minutes to remove organic impurity
	of aluminum sheets. Cleaned aluminium sheets were electro-polished using electrolyte
	with 1:4 mixture of perchloric acid (HCLO4) and ethanol at constant DC voltage 25V
	and current of 3 Ampere for two minutes and the polished sheet was washed many times using
	DI water and dried in air. Before anodization, Hexaphosphonic acid (HPA) coating was
	done at interface of polished and unpolished region of sheet to avoid the surface impurity
	caused by surface current. oxalic acid solution (0.15 M) was used as electrolyte.
	Alumina films were prepared by two step anodization. First step anodization was done
	for 14 hours at 40V and 50 V depending on pore diameters. Oxidized alumina during
	first anodization were removed by dipping anodized sheet in Cromic acid solution(
	2gram cromic oxide (Cr2O3) in 200ml water) for 8 hours. After that, sheet was washed
	many times with DI water. Again HPA coating was done at interface of polished and
	unpolished region and second anodization was done for two hour . Thickness of film
	depends of time of second anodization. Finally , previously coated HPA was removed
	by dipping sheet in toluene and cleaned with acetone and DI water.Anodized aluminium sheet was mounted in teflon jacket and bottom part of sheet
	was supported by coverslip glass. Top part of alumina was remove by sodium hydroxide (
	NaOH) solution in water (0.02mg/ml) and kept for 15 minutes. Finally, aluminium
	metal layer was removed by solution of copper chloride (CuCl2) and hydrochloric acid (HCl). After removing the metal layer , residual thin alumina membrane  was gently
	removed from jacket and washed with DI water many times and kept for drying. for removing the bottom barrier layer, top surface of membrane was covered with coverslip
	glass and membrane  was dipped in 6$\%$H3PO4 (ortho Phosphoric acid for 40 min
	and membrane was again cleaned with DI water and kept drying.
	silver nano wire was grown in porous alumina membrane using DC electro deposition
	technique. electrolyte used for electro deposition contains the Silver Bromide (AgBr) (0.1M), sodium hyposulphite (Na2SO3), Sodium Sulphite (NaSO3). A thin
	layer of Gold was sputtered on top surface of membrane to cover the pore completely which
	will act as nano electrode in electrodeposition. Three electrode system was used in
	electrodepsition: Graphite sheet as Anode, a (HgCl2=AgCl) referece electrode to measure
	the electrolyte potential near membrane, gold coated alumina membrane act as
	cathode. Electro deposition of silver nano wire in porous alumina membrane was done
	at constant voltage -700mV for 1 hours. Membrane was cleaned with DI water several time after electro-depsition . Top unfilled alumina was removed by dilute solution of
	NaOH (1mg/ml). Previously deposited gold was cleared using focus ion beam (FIB) technique\cite{indukuri2017broadband,kanungo2010experimental}.
	\section{Sample preparation method of QDs/HMM templates}
	Finally FIB eteched sample was transfer on RCA cleaned coverslip using toluene drop. Polymer layer of PS was spin coated on HMM using spin coater with speed 3000 RPM for 60
	second.Thickness of polymer layer was controlled by concentration of PS solution. PS solution with concentration 1mg/ml gives 5nm thickness for above mentioned parameter of spin
	coater.
	CdSe QD monolayer was transferred on HMM using Langmuir-Schaefer (LS) deposition
	techniqueLS method. In the LS deposition technique, a hydrophobic CdSe QD solution in
	chloroform (0.6mg/ml) was spread drop by drop on a water surface of a Langmuir-Blodgett
	(LB) trough until pressure reaches saturation level. Finally, the QD monolayer on the water
	surface is compressed by barriers till trough area 50x50$\text{cm}^{2}$ to surface pressure 30mN/m
	which is the optimal pressure for compact QD monolayer formation. The QD monolayer is
	transferred onto HMM by touching the HMM to floating QD monolayer on water surface of
	LB trough.
	\section{Experimental set up}
	An inverted microscope is used for capturing the Fourier images for back focal plane (BFP) imaging. QD on HMM templates was excited by 633nm He-Ne laser by focusing 100X (NA= 0.95) objective and emission was collected using the same objective lens. A Pinhole was placed in conjugate image plane of lens L4 and it acts as spatial filter. Bertrand lens L5 transform a real image created by L4 in to Fourier image and it is focused on EM CCD to record the Fourier images.
	The laser line was rejected by a combination of 633nm edge and notch filters. Collected emission was projected on Fourier plane and analysed using EM CCD to get Fourier image as shown in Figure S2. Fourier image was obtained by imaging the BFP of objective, which provides the information about the wave vector of emitted light from focused excitation spot on QDs/HMM sample. Image view size of Fourier image was limited by numerical aperture of objective. 
	Circular polarized (CP) excitation is generated by a combination of half wave plate and quarter wave plate. The helicity of CP was switched from Left-handed Circular Polarization (LCP) to Right-handed Circular Polarization (RCP) by changing the angular orientation of quarter wave plate with respect to incident polarization direction. Emission from QDs on HMM templates was collected for
	different incident excitation polarisation:  unpolarised, LCP and RCP.
	\begin{figure*}[t]
		\centering
		\includegraphics[scale=.7]{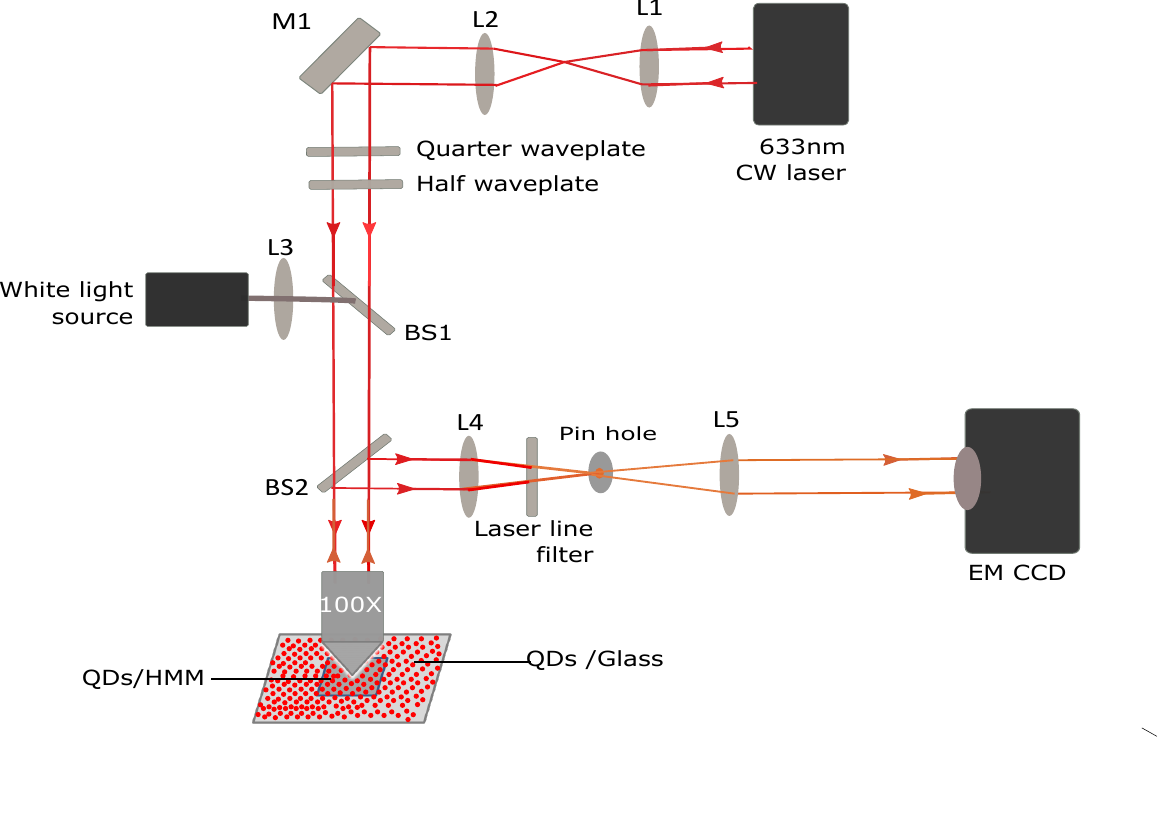}
		\caption{shows ray diagram of optical set up for Fourier imaging. }
		\label{fig:fl_map}
	\end{figure*}

		
	\section{Spacer thickness ($d$) dependent Photolumenescence(PL) spectra}
	Far-field PL spectra of QDs on HMM templates was measured using "WITec Alpha 300" system. QD monolayesr on HMM and on glass was excited by focused 514 nm unpolarised laser using 20X (NA=0.35)objective and PL  emission spectra was collected by same objective and laser line was rejected by 520 nm edge filter\cite{indukuri2017broadband}. 
	Figure S3 shows distance dependent far-field PL spectra for $d$=10 nm and 20 nm from QDs on HMM sample and $d$=10 nm for QDs on glass, which clearly shows splitting for smaller $d$ = 10 nm and dissappears for larger $d$ =20 nm. The spectra on glass is largely independent of $d$.
		\begin{figure}[h!]
		\centering
		\includegraphics[scale=0.60]{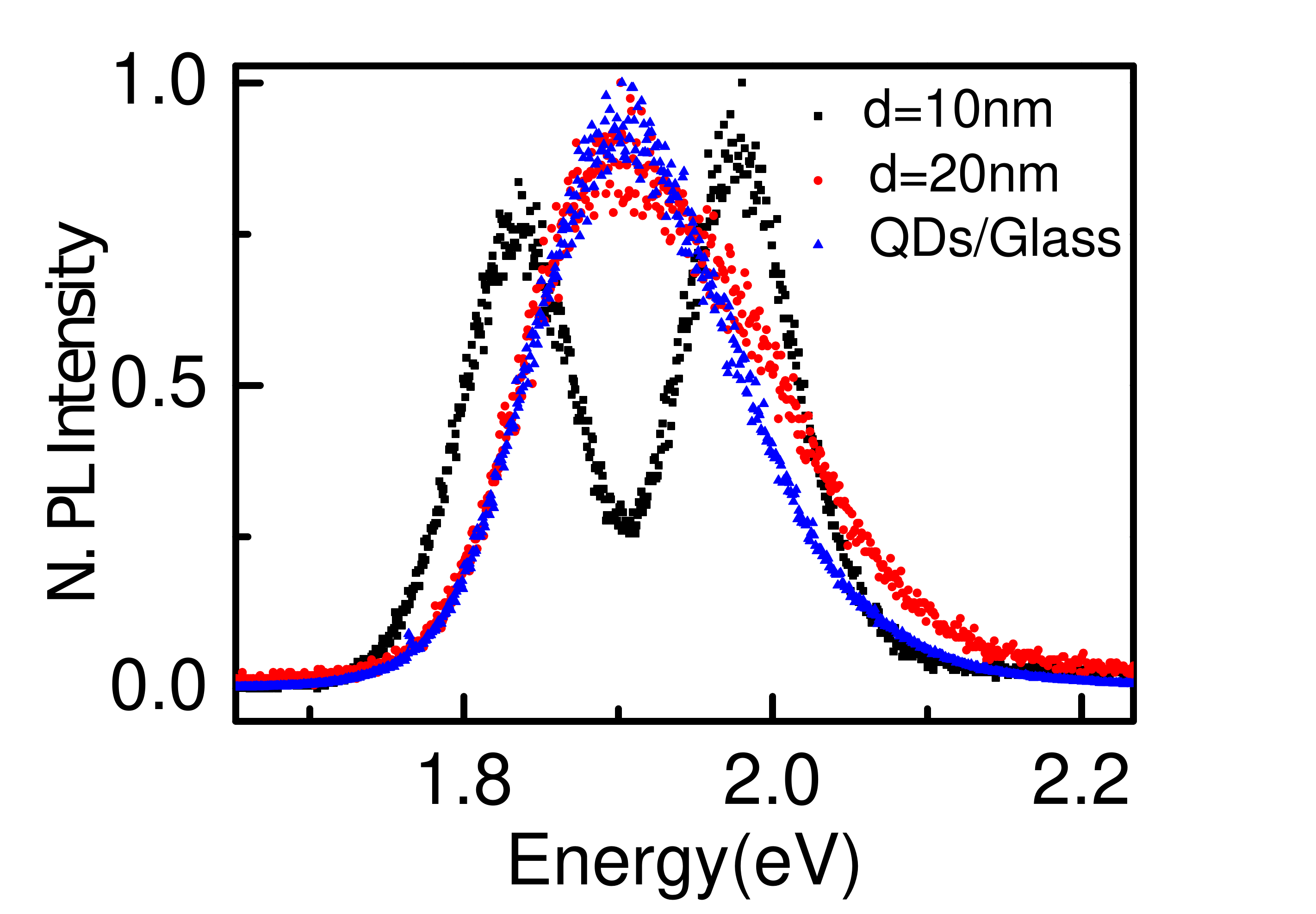}
		\caption{Far-field PL spectra of QDs on HMM
			for spacer thickness $d$=10nm, $d$=20nm and for $d$=10nm for QDs on glass.}
	\end{figure}
	\section{Coupling dependent directional emission }
	Figure S4 shows Fourier image of QDs on HMM which suggest  that the directionality  effect seen for $d$=10nm sample disappears. This suggests a connection between splitting in Fig S3 and the directionality. 
	
	\begin{figure}[h!]
		\centering
		\includegraphics[scale=0.7]{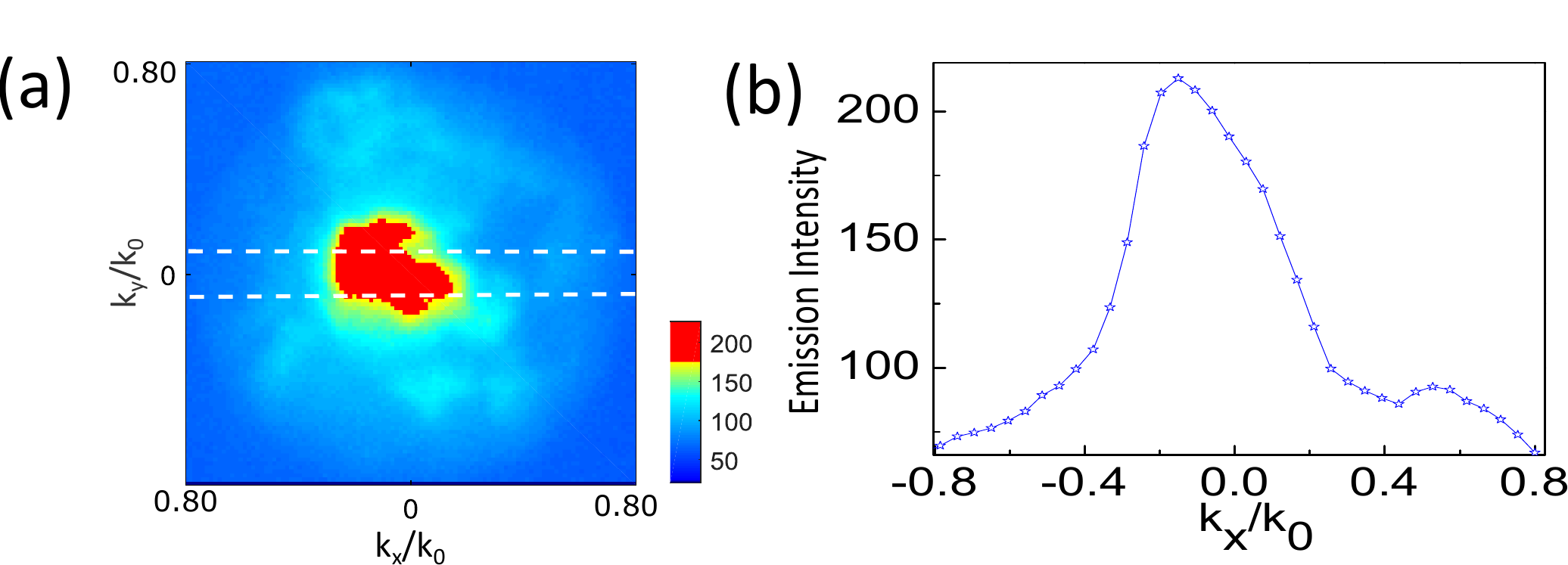}
		\caption{(a) shows Fourier  image of QD coupled HMM
			emission with unpolarised excitation for spacer thickness $d$=20nm.(b)Shows line profile of  QD emission intensity as a function of in-plane wave vector $k_{x}/k_{0}$ for the patterns shown in (a).}
	\end{figure}
	
		\section{Effective medium theory for nanowire HMMs}
	In experiment, anisotropic HMMs are composed of an array of sliver nanowires in the $\text{Al}_{2}\text{O}_{3}$ dielectric host. We assume that the single sliver nanowire has a cross-sectional area $a$ and the array unit has a cross-sectional area $A$. The filling fraction factor can be defined as $f=a/A$. Here, we set $f=0.15$. The relative dielectric constant of the $\text{Al}_{2}\text{O}_{3}$ dielectric host is taken as $\epsilon_{d}=2.56$ and independent of wavelength. These values ($\epsilon_{m}$) of sliver can be taken from  Ref.~[1]. By using the Maxwell-Garnet theory, the effective medium dielectric constants for the HMMs can be expressed as 
	\begin{subequations}  
		\begin{align}
		\epsilon_{\perp}=&f \epsilon_{m}+(1-f ) \epsilon_{d}\tag{S1a},\\ 
		\epsilon_{\parallel}=&\frac{(1+f)\epsilon_{m}\epsilon_{d}+(1-f)\epsilon_{d}^{2}}{ (1-f)\epsilon_{m}+(1+f)\epsilon_{d}}\tag{S1b}, 
		\end{align}
	\end{subequations}
	where $\epsilon_{z}=\epsilon_{\perp}$ is the permittivity perpendicular to the optical axis, and $\epsilon_{x}=\epsilon_{y}=\epsilon_{\parallel}$ is the permittivity parallel to the optical axis in the x-y plane.
	\begin{figure}
		\centering
		\includegraphics[scale=0.8]{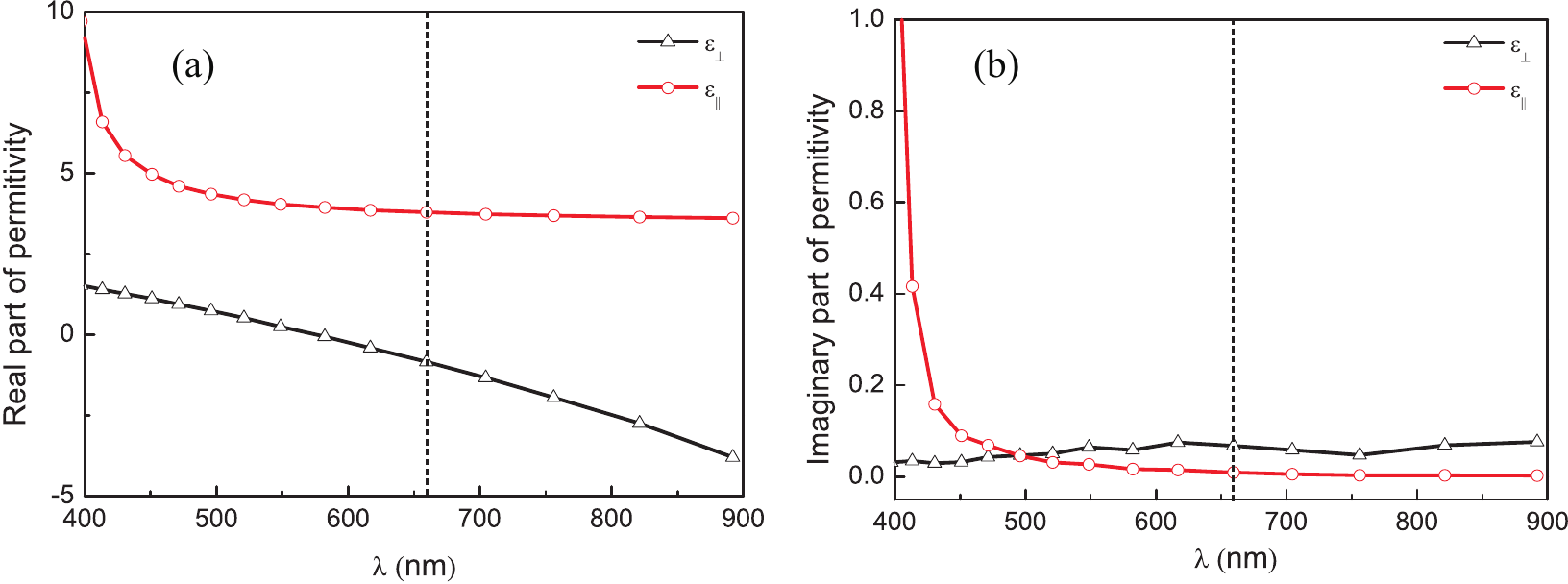}
		\caption{Real part (a) and imaginary part (b) of the dielectric constants for the HMMs as a function of wavelength by using the effective medium theory.}
		\label{fig:fl_map}
	\end{figure}
	
	Different diagonal components of the dielectric constant tensor for the HMMs as a function of wavelength are plotted in Fig.~S5. Neglecting dissipation for a moment, we can find for some frequency range $\epsilon_{\parallel}>0$ and $\epsilon_{\perp}<0$, which satisfies the condition  $\epsilon_{\parallel}\epsilon_{\perp}<0$ for HMMs. By using linear fitting,  for incident wavelength $\lambda=660~\text{nm}$ the permittivity for the HMMs is $\epsilon_{\parallel}=3.79+0.0088i$  and $\epsilon_{\perp}=-0.843+0.0672i$ from Fig.~S5 (corresponding to the black dotted lines). 
	 Moreover, the $z$ components of the wave vector for the s-polarized and p-polarized modes in the HMMs are given by $k_{3z}^{s}=\sqrt{k_{0}^2\epsilon_{\parallel}-\kappa^2}$ and $	k_{3z}^{p}=\sqrt{k_{0}^2\epsilon_{\parallel}-\kappa^2\frac{\epsilon_{\parallel}}{\epsilon_{\perp}}}$, respectively.
	
	\section{Theoretical modeling: field distribution by using Green's function}
	\begin{figure}[h!]
		\centering
		\includegraphics[width=7cm]{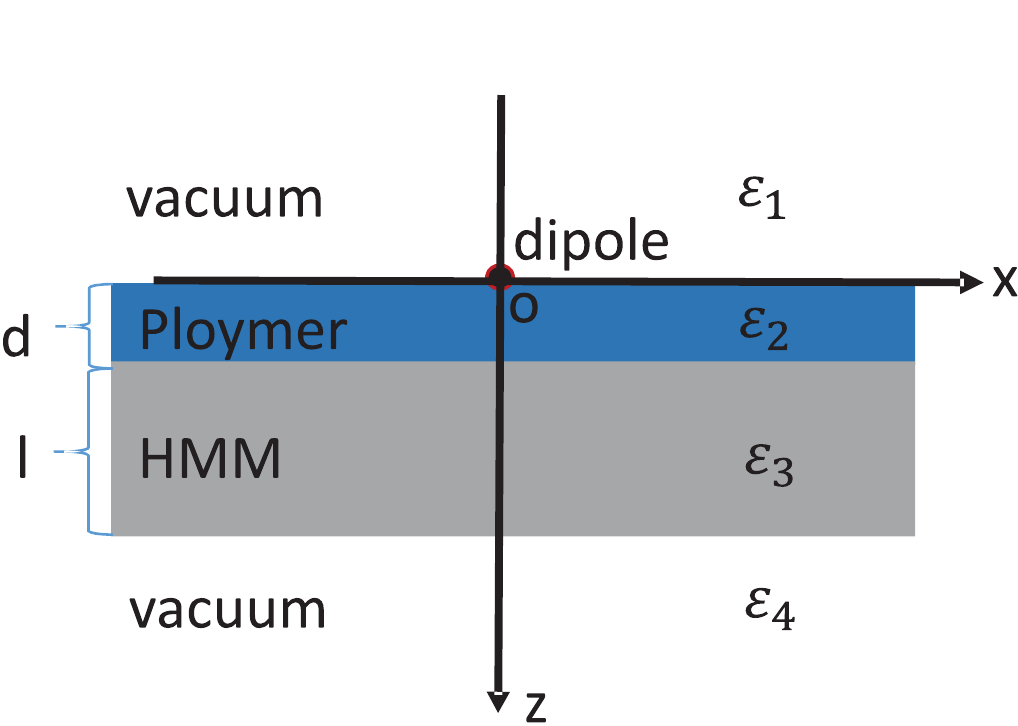}
		\caption{(Color online)  Schematic of a dipole embedded into a three-dimensional multilayer nanostructure. The relative dielectric constant in the region $i$ is $\epsilon_{i}$.}
	\end{figure}
	We next develop a model for the observations. By using the rigorous solutions of Maxwell equations, we can obtain the electric field in the HMM induced by an electric dipole. Unlike the layered medium in the previous work, the HMM in the present work is different and is characterized by complex effective dielectric tensor. We consider an electric dipole radiating in a multilayer nanostructure, as shown in Fig.~S6. In the
	experiment, compact monolayers of quantum dots, which is placed at a polymer spacing $d$ to the surface of hyperbolic metamaterial templates, can be treated as a local in plane electric dipole. The hyperbolic metamaterial templates are designed by utilizing sliver metallic
	nanowires in a $\text{Al}_{2} \text{O}_{3}$ dielectric host with a thickness $l = 800~\text{nm}$. The materials are nonmagnetic with permeability $\mu=1$. The relative dielectric constant in the region $j$ is $\epsilon_{j}$ ($j=1, 2, 3, 4$). We assume the dipole at a position $\textbf{r}_{0}$ in the region 1 has a dipolar moment $\textbf{p}={p}\hat{\textbf{p}}_{0}$, where $\hat{\textbf{p}}_{0}$ is the unit vector denoting the polarized direction. Then, the solution of the electric field in this multilayer nanostructure is equal to a one-dimensional problem.
	
	The electric field at a position $\textbf{r}$ in the region 1 is related to the primary Green function and the reflected Green function with
	\begin{equation} 
	\textbf{E}(\textbf{r},\textbf{r}_{0};\omega)= \frac{\epsilon_{1}\omega^{2}}{c^{2}} [\overleftrightarrow{\textbf{G}}_{0} (\textbf{r},\textbf{r}_{0};\omega) +\overleftrightarrow{\textbf{G}}_\text{R} (\textbf{r},\textbf{r}_{0};\omega)]\cdot \textbf{p},
	\end{equation}
	where $\omega$ is the frequency of the dipole, and $c$ is the velocity of light in vacuum. The dyadic function $\overleftrightarrow{\textbf{G}_{0}}(\textbf{r},\textbf{r}_{0};\omega)$ in an isotropic medium is the solution of
	\begin{equation} 
	[\nabla \times \nabla \times- \epsilon_{1} \frac{\omega^{2}}{c^{2}}\overleftrightarrow{\textbf{I}}\cdot]\overleftrightarrow{\textbf{G}_{0}} (\textbf{r},\textbf{r}_{0};\omega) =\overleftrightarrow{\textbf{I}} \delta(\textbf{r},\textbf{r}_{0}),
	\end{equation}
	where $\overleftrightarrow{\textbf{I}}=\hat{\textbf{x}}\hat{\textbf{x}}+\hat{\textbf{y}}\hat{\textbf{y}}+\hat{\textbf{z}}\hat{\textbf{z}}$ is the unit dyadic. Using the Weyl's expansion, the dipole Green function $\overleftrightarrow{\textbf{G}_{0}}(\textbf{r},\textbf{r}_{0};\omega)$ in free space can be decomposed of plane waves with different angular spectrum. Each plane wave is incident to the interface and then reflected or transmitted. In the region 1 the Green function $\overleftrightarrow{\textbf{G}}_\text{R} (\textbf{r},\textbf{r}_{0};\omega)$ is consisting of all the reflected waves. We assume that the multilayered structure considered above is strictly planar so that $p$-polarized (TM) waves and $s$-polarized (TE) waves don't interact with each other.
	Here we are more concerned about the electric field at the position $\textbf{r}_{0}$, therefore the contribution from the term $\overleftrightarrow{\textbf{G}_{0}}(\textbf{r}_{0},\textbf{r}_{0};\omega)$ can be neglected as this contribution is not connected with the excitation of high $k$ vectors in HMM.  In the Fourier transform space, the reflected Green’s function is given by
	\begin{equation}
	\begin{split}
	\overleftrightarrow{\textbf{G}}_\text{R} (\textbf{k},\omega;z,z_{0})=& \frac{i}{2} (\overleftrightarrow{\textbf{M}}^\text{s} +\overleftrightarrow{\textbf{M}}^\text{p})
	e^{-i k_{1z}(z+z_{0})},
	\end{split}
	\end{equation}
	where
	\begin{subequations} 
		\begin{align}
		\overleftrightarrow{\textbf{M}}^\text{s}&=\frac{r^\text{s}(k_{x},k_{y})}{k_{1z}(k_{x}^2+k_{y}^2)}
		\left(
		\begin{array}{ccc}
		k_{y}^2 & -k_{x}k_{y} & 0\\
		-k_{x}k_{y} & k_{x}^2 & 0\\
		0 & 0 & 0\\
		\end{array}
		\right),
		\\
		\overleftrightarrow{\textbf{M}}^\text{p}&=\frac{-r^\text{p}(k_{x},k_{y})}{k_{1}^{2}(k_{x}^2+k_{y}^2)}
		\left(
		\begin{array}{ccc}
		k_{x}^{2} k_{1z}& k_{x}k_{y}k_{1z} & k_{x}(k_{x}^{2}+k_{y}^{2})\\
		k_{x}k_{y}k_{1z} & k_{y}^{2} k_{1z} & k_{y}(k_{x}^{2}+k_{y}^{2})\\
		-k_{x}(k_{x}^{2}+k_{y}^{2}) & -k_{y}(k_{x}^{2}+k_{y}^{2}) & -(k_{x}^{2}+k_{y}^{2})^{2}/k_{1z}\\
		\end{array}
		\right).
		\end{align}
	\end{subequations}
	Considering $z=z_{0}=0$, one can find the reflected electric field can be written as
	\begin{equation}
	\textbf{E}_\text{R}(\textbf{k},\omega)= \frac{i\epsilon_{1}\omega^{2}p}{2c^{2}}   (\overleftrightarrow{\textbf{M}}^\text{s} +\overleftrightarrow{\textbf{M}}^\text{p})\cdot \textbf{p}_{0}.
	\end{equation}
	In these equations, $k_{x}$ ($k_{y}$) is the $x$ ($y$) component of the wave vector, and $k_{1z}$ and is the $z$ component of the wave vectors in vacuum ($k_{1}=\omega/c$), respectively. Note that, $r^{s}$ and $r^{p}$ are reflection coefficients of s-polarized and p-polarized waves, respectively.
	We assume in a two dimensional space the unit dipole moment is $ \textbf{p}_\text{0}=p_{x} \hat{\textbf{x}}+p_{z} \hat{\textbf{z}}$ with $p_{x}^{2}+p_{z}^2=1$, the electric intensities can be expressed as
	\begin{equation} 
	|\textbf{E}_\text{R}(\textbf{k},\omega)|^{2}=\frac{\omega^{2}p^{2}}{4c^2} \cdot I,
	\end{equation}
	where
	\begin{equation} 
	I=k_{1}^2[|(M^{s}_{xx}+M^{p}_{xx})p_{x}+M^{p}_{xz}p_{z}|^{2}+|M^{p}_{zx}p_{x}+M^{p}_{zz}p_{z}|^{2}].
	\end{equation}
	It's easy to find that $I$ is an even function of the wave vector component $k_{y}$. Note that the Eq.~S(8) includes the contribution of the propagating ($\kappa<k_{0}=\omega/c$) and evanescent ($\kappa>k_{0}$) waves. By comparison to the case with absorption in sliver in the main body, we also give the electric intensity maps as functions of the wave vector components ($k_{x}$ and $k_{y}$) in Fig.~S7. The components of permittivity tensor are given by $\epsilon_{\perp}=-0.8769$ and $\epsilon_{\parallel}=2.5974$ for the emission wavelength $\lambda=660~\text{nm}$. Similar results can be obtained in Fig.~S7 compared to these in Fig.~4. However, the bandwidth of the frequency spectra is ultra-narrow at $k_{x}=\pm1.1964k_{0}$, and the chiral parameter is $C=0.845$ in Fig.~S7(c). 
	
	\begin{figure*}
		\centering
		\includegraphics[scale=0.65]{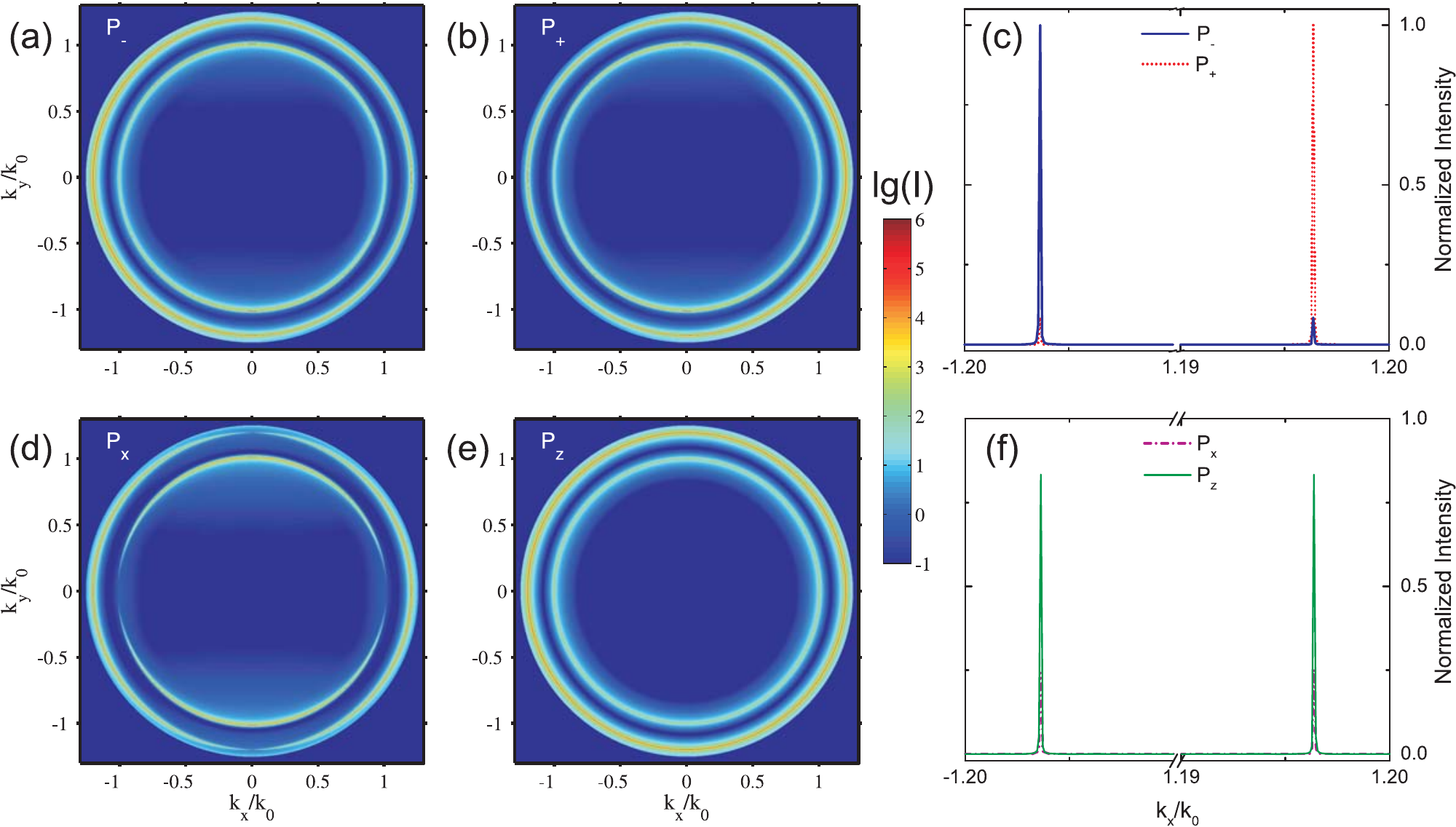}
		\caption{(Color online) Electric field intensity maps when neglecting the absoption in sliver with different polarization, (a) $ \textbf{P}_\text{-}=\frac{1}{\sqrt{2}}( \hat{\textbf{x}}-i\hat{\textbf{z}})$, (b) $ \textbf{P}_\text{+}=\frac{1}{\sqrt{2}}( \hat{\textbf{x}}+i\hat{\textbf{z}})$, (d) $ \textbf{P}_\text{x}= \hat{\textbf{x}}$, and (e)$\textbf{P}_\text{z}= \hat{\textbf{z}}$.
			Intensity profile induced by circularly polarized (c) and linearly polarized (f) dipoles as a function of $k_{x}$ are plotted for $k_{y}=0$. The electric field $E_{c}$ used for normalization corresponds to the larger peak value induced by a circularly polarized dipole in  Fig.~S7(c).}
	\end{figure*}
	
	\section{The reflection coefficients of the multilayered structure}
	Bellow we will derive the reflection coefficients in the multilayer structure. Consider a p-polarized electromagnetic wave first, the only magnetic field is along the $\hat{\textbf{y}}$ direction and conserved in the $\hat{\textbf{x}}-\hat{\textbf{z}}$ plane. The field in each layer can be described as the superposition of upward and downward propagating waves due to the reflection at the layer's boundaries. Therefore, the magnetic field component in the multilayer structure is given by
	\begin{equation}
	H_{y}=e^{ik_{x}x}
	\left\{
	\begin{split}
	&e^{ik_{1z}z}+r^{p}e^{-ik_{1z}z}\quad  &(z<0)\\
	&t_{2}e^{ik_{2z}z}+r_{2}e^{-ik_{2z}(z-d)}\quad &(0\leq z<d)\\
	&t_{3}e^{ik_{3z}(z-d)}+r_{3}e^{-ik_{3z}(z-d-l)} \quad &(d\leq z<d+l)\\
	&t_{4}e^{-ik_{4z}(z-d-l)} \quad &(z\geq d+l)
	\end{split}
	\right.
	\end{equation}
	In Eq.~(S9) we have neglected the term $\text{exp}(-i\omega t+i k_{y}y)$. The z component of the wave vector in the $j$th layer is $k_{jz}$ ($j=1, 2, 3, 4$), $r_{j}$ and $t_{j}$ are the reflection and transmission coefficients of the $j$th layer (j=2, 3, 4), respectively. 
	
	The electric field can be solved by the Maxwell's curl equation, $-i\omega \epsilon_{j}\textbf{{E}}=\bigtriangledown \times \textbf{{H}}$. Then by applying the boundary conditions [$\textbf{{n}}\times (\textbf{H}_{i+1}-\textbf{H}_{i})=0$, and $\textbf{{n}}\cdot (\epsilon_{i+1}\textbf{E}_{i+1}-\epsilon_{i}\textbf{E}_{i})=0$],
	we can obtain six linear equations consisting of unknown reflection and transmission coefficients. For the y-component of magnetic field ($H_{y}$), we obtain
	\begin{subequations}  
		\begin{align}
		1+r^{p}&=t_{2}+r_{2}e^{ik_{2z}d},\\
		t_{2}e^{ik_{2z}d}+r_{2}&=t_{3}+r_{3}e^{ik_{3z}l},\\
		t_{3}e^{ik_{3z}l}+r_{3}&=t_{4}, 
		\end{align}
	\end{subequations}
	and for the x-component of electric field ($E_{x}$), we can get
	\begin{subequations}  
		\begin{align}
		1-r^{p}&=D_{12}^{p}(t_{2}-r_{2}e^{ik_{2z}d}), \\
		t_{2}e^{ik_{2z}d}-r_{2}&=D_{23}^{p}(t_{3}-r_{3}e^{ik_{3z}l}),\\
		t_{3}e^{ik_{3z}l}-r_{3}&=D_{34}^{p}t_{4},
		\end{align}
	\end{subequations}
	where $D_{ij}^{p}=\frac{\epsilon_{i}k_{jz}}{\epsilon_{j}k_{iz}}$. 
	The Fresnel reflection coefficient is defined as  $R_{ij}^{p}=\frac{1-D_{ij}^{p}}{1+D_{ij}^{p}}$. By solving the above equations, the effective reflection coefficient $r^{p}$ is given by
	\begin{equation} 
	\begin{split}
	{r}^{p}=\frac{{R}_{12}^{p}+\tilde{R}_{23}^{p}e^{2ik_{2z}d}}{1+{R}_{12}^{p}\tilde{R}_{23}^{p}e^{2ik_{2z}d}},\\
	\tilde{R}_{23}^{p}=\frac{{R}_{23}^{p}+{R}_{34}^{p}e^{2ik_{3z}l}}{1+{R}_{23}^{p}{R}_{34}^{p}e^{2ik_{3z}l}}.
	\end{split}
	\end{equation}
	Furthermore, for the s-polarized wave, the y-component of the electric field ($E_{y}$) has the same formulations as given in Eq.~(S9). The similar results for the reflection and transmission coefficients in each layer are obtained but with $D_{ij}^{s}=\frac{k_{jz}}{k_{iz}}$,and $R_{ij}^{s}=\frac{1-D_{ij}^{s}}{1+D_{ij}^{s}}$. Therefore, the electric field in each layer including the contributions of the TM and TE waves can be obtained.
	
	\begin{figure}
		\centering
		\includegraphics[scale=0.7]{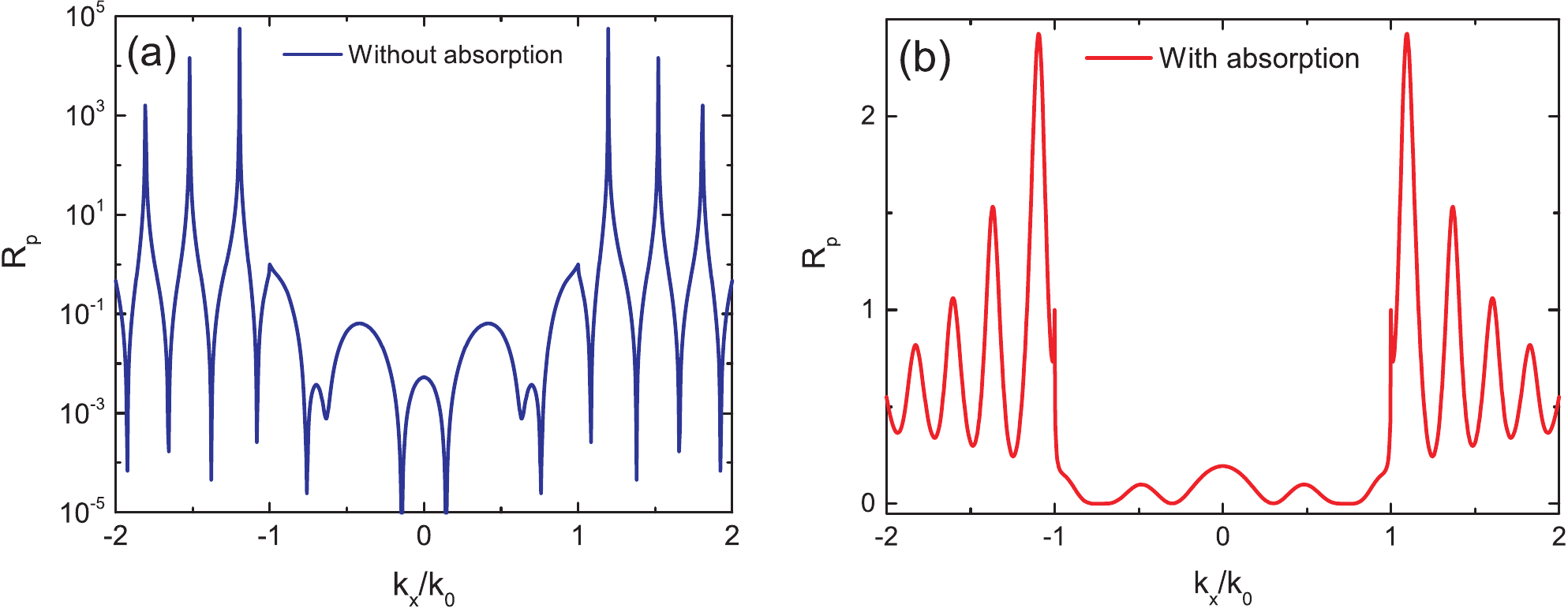}
		\caption{ The reflection $R_{p}=|r^{p}|^2$ of p-polarized waves versus the wave vector component $k_{x}$ for $k_{y} = 0$ and $\lambda=660~\text{nm}$ when the hyperbolic metamaterial is (a) without absorption, (b) with absorption.}
	\end{figure}
	
	The reflection ($R_{p}=|r^{p}|^2$) of the TM waves across the HMMs is plotted in Fig.~S8 for $k_{y}=0$. The dielectric constants and geometrical parameters corresponding to the experiment are $\epsilon_{1}=\epsilon_{4}=1$, $\epsilon_{2}=2.5$, $d=10~\text{nm}$, and $l=800~\text{nm}$. As discussed above, for the type-$\text{\uppercase\expandafter{\romannumeral1}}$ HMMs only the perpendicular componet of the dielectric tensor is negative. In Fig.~S8 the reflection is symmetric along the axis $k_{x}=0$, which means the multilayered nanostructure supports bidirectional propagating eigenmodes. For the propagating waves ($k_{x}<k_{0}$)
	the reflection is much smaller than $1$. Moreover, we note that multiple peaks present in the reflection of large wavevector waves ($k_{x}>k_{0}$) are evanescent and corresponding to different coupled surface modes in HMMs. In Fig.~S8(a) when neglecting the loss in metal, the reflections at peaks $k_{x}/k_{0}=\pm 1.805, \pm 1.52, \text{and} \pm 1.196$ can be larger than $10^{3}$. However, the case is different when considering the absorption. The maximum of the reflection at $k_{x}/k_{0}=\pm 1.096$ is about 2.5. The difference is due to the wave vector component $k_{3z}^{p}$, which is a real number without absorption. For some $k_{x}$ values, the denominator in $r^{p}$ will become a pure small imaginary number. However, when considering the loss in metal, $k_{3z}^{p}$ is a complex number. In the text we consider the contribution of the first evanescent mode to the electric field.

	%
	
